# Reimagining Heliophysics: A bold new vision for the next decade and beyond

Ian J. Cohen[1], Dan Baker[2], Jacob Bortnik[3], Pontus Brandt[1], Jim Burch[4], Amir Caspi[4], George Clark[1], Ofer Cohen[5], Craig DeForest[4], Gordon Emslie[6,7], Matina Gkioulidou[1], Alexa Halford[8], Aleida Higginson[8], Allison Jaynes[9], Kristopher Klein[10], Craig Kletzing[9], Ryan McGranaghan[11], David Miles[9], Romina Nikoukar[1], Katariina Nykyri[12], Larry Paxton[1], Louise Prockter[1], Harlan Spence[13], William H. Swartz[1], Drew L. Turner[1], Joe Westlake[1], Phyllis Whittlesey[14], Michael Wiltberger[15]

[1] *JHU/APL*
[2] *University of Colorado Boulder/Laboratory for Atmospheric and Space Physics*
[3] *UCLA*
[4] *SwRI*
[5] *University of Massachusetts, Lowell*
[6] *Western Kentucky University*
[7] *University of Alabama at Huntsville*
[8] *NASA Goddard Space Flight Center*
[9] *University of Iowa*
[10] *University of Arizona*
[11] *Orion Space Solutions*
[12] *Embry-Riddle Aeronautical University*
[13] *UNH*
[14] *UC Berkeley/SSL*
[15] *NCAR/HAO*

## Summary

The field of Heliophysics has a branding problem. We need an answer to the question: "*What is Heliophysics?*", the answer to which should clearly and succinctly defines our science in a compelling way that simultaneously introduces a sense of wonder and exploration into our science and our missions. Unfortunately, recent overreliance on space weather to define our field, as opposed to simply using it as a practical and relatable example of applied Heliophysics science, narrows the scope of what solar and space physics is and diminishes its fundamental importance. Moving forward, our community needs to be bold and unabashed in our definition of Heliophysics and its big questions. We should emphasize the general and fundamental importance and excitement of our science with a new mindset that generalizes and expands the definition of Heliophysics to include new "frontiers" of increasing interest to the community. Heliophysics should be unbound from its current confinement to the Sun-Earth connection and expanded to studies of the fundamental nature of space plasma physics across the solar system and greater cosmos. Finally, we need to come together as a community to advance our science by envisioning, prioritizing, and supporting – *with a unified voice* – a set of bold new missions that target compelling science questions– even if they do not explore the traditional Sun- and Earth-centric aspects of Heliophysics science. Such new, large missions to expand the frontiers and scope of Heliophysics science large missions can be the key to galvanizing the public and policymakers to support the overall Heliophysics program.

*What IS Heliophysics?*

The field of Heliophysics has a branding problem. We, as a community, have failed to clearly and succinctly define our science in a way that underscores its fundamental and universal relevance while also making a pitch that is compelling to the general public. This has hurt not only our visibility and awareness in the public zeitgeist but also our ability and effectiveness in appealing to policymakers in Congress and the Executive Branch. Together, the lack of "brand recognition" for Heliophysics and our limited funding inhibit our ability to maximize our potential for understanding solar and space physics as well as to inspire, recruit, and train the next generation of solar and space physics researchers at the very time when the government, commercial, and societal demand to understand the space environment is increasing exponentially.

**When people ask "*What is Heliophysics?*", the answer should speak to the fundamental understanding of the very nature of near-planetary, interplanetary, interstellar, and intergalactic space itself (i.e., plasma - by far the most common state of matter in the universe). It should encompass, in both name and practice, stellar-planetary and moon-magnetosphere interactions across the solar system, galaxy, and cosmos.** By its very nature, involves a closely coupled combination of a variety of scientific elements, including astrophysics, space plasma physics, atmospheric physics, and geophysics, requiring extraordinarily detailed measurements obtained both through remote sensing and multi-scale in-situ measurements to study the only star, planetary space environments, and astrosphere to which we have direct access. It is also the comparative study of planetary magnetospheres and atmospheres across the solar system and beyond.

Heliophysics is and will continue to be many exciting things. Today, it is the amazing photographs of auroral displays seen from the ground and the International Space Station. It is the discovery of the mysterious "STEVE" (Strong Thermal Emission Velocity Enhancement) emissions in the sky by amateur citizen scientists. It is using large-scale advanced ground-based solar telescopes, like NSF's Daniel K. Inouye Solar Telescope (DKIST) to probe the detailed dynamics of the outer regions of the Sun. It is the hypnotizingly complex videos of the Sun's boiling plasma surface and corona revealed by NASA's Solar Dynamics Observatory (SDO). It is exploring the nature of the solar wind and the very edges of the solar system with NASA's Interstellar Mapping and Acceleration Probe (IMAP). It is deploying four Magnetospheric Multiscale (MMS) spacecraft that have unlocked the mysteries of magnetic reconnection at the electron scale. It is developing advanced computer simulations of the environments in near-Earth space, the Sun, and interplanetary and interstellar space. And it is launching a spacecraft like NASA's Parker Solar Probe to touch our Sun.

Despite these successes, Heliophysics has been increasingly pigeonholed over the last decade – at least in the eyes of the public and policymakers – into only being important as it relates to "space weather" (i.e., conditions and variability of the natural space environment that result in hazardous effects on – and presenting increased risk of damage to – human systems). When competing against the easily digestible and exciting distilled themes of other disciplines, e.g., Planetary Science ("The search for life"), Earth Science ("Understanding our home planet"), and Astrophysics ("Exploring the origin and nature of the universe") – Heliophysics has historically struggled. Often, selling Heliophysics to those outside the field is a challenge even though stellar and space plasmas and their processes represent some of the most genuinely universal and fundamentally important physics in the cosmos. The Heliophysics community has been hard-pressed to distill our major fundamental questions into exciting and intuitive problems that capture the public's attention. As such, we have understandably reverted to a very practical, applied definition – i.e., space weather – to justify our discipline as we advocate for our science and appeal for funding.

Space weather is a powerful practical example and something that brings home the everyday relevance of our field when speaking to the public or policymakers. Space weather has a major impact on policy both in the U.S. and worldwide. Many U.S. Congressional committees deal with issues for which space physics is relevant, including those for Homeland Security, Armed Services, Intelligence; Energy; Science; Transportation; Agriculture; Commerce; and Education.[3] Space weather is an appropriate and viable lens through which to frame the relevance of solar and space plasma physics and much success has been made over the last few years in the organization and codification of a federal strategy regarding space weather through the National Space Weather Action Plan and the PROSWIFT legislation. **However, overreliance on space weather to define *our field, as opposed to simply using it as a practical and relatable tool and valuable example of applied Heliophysics science, narrows the scope of what solar and space physics is and diminishes its fundamental importance*.** We must realize and delicately balance the temptation to use space weather to define our science based only on the increasing societal relevance of space weather effects. Especially in light of humanity's ambition to expand into the solar system through even greater emphasis on human exploration (i.e., desires to establish a crewed lunar station, colonize Mars, and mine the asteroid belt), it will become increasingly critical to understand and consider the nature of the space environment throughout the solar system, but not to *DEFINE* our all heliophysics by that application.

**Moving forward, hopefully emboldened by the next Decadal Survey, the solar and space physics community should focus emphasis on the general and fundamental importance and excitement of its science with a new mindset that generalizes and expands the definition of Heliophysics to include new "frontiers" of increasing interest to the community:**

> *Heliophysics is a fundamental science discipline that is the study of the very nature of plasmas throughout space, originating with our own Sun and heliosphere and extending to planetary atmospheres and magnetospheres, stellar atmospheres and astrospheres, interstellar space, and more. It is interwoven with planetary science, astrophysics, and geoscience.*

Such a fundamental redefinition of Heliophysics and a commensurately ambitious Heliophysics Program of science missions and discovery will require a correspondingly ambitious budget, coupled with strong support within NASA, the White House Office of Science and Technology Policy, the Office of Management and Budget, and Congress. To counteract the unfortunate recent trend in Heliophysics funding and enable new, innovative, and inspiring solar and space plasma physics missions throughout the solar system and into local interstellar space, the Heliophysics community needs to be actively engaged in advocacy for our field. **Armed with bold ideas and exciting ambitions, inspiring stories, impressive visuals, and accessible language, Heliophysics researchers need to unapologetically promote their fundamental and practical research and the genuinely universal scope of Heliophysics to the public and policymakers.** Celebrating research successes is a necessary public service in the best interest of the Heliophysics community overall.

*A Bold Decadal Vision*

**Looking forward to the next decade, our scientific community needs to be bold and unabashed in our definition of Heliophysics and its big questions.** Unfortunately, despite the fundamental nature of solar and space plasma physics and space weather's importance to human society and the national interests of the United States, Heliophysics receives the lowest funding amongst the primary space science Divisions (Planetary Science, Earth Science, Heliophysics, Astrophysics) within the NASA Science Mission Directorate and is at constant risk of further budget reductions.

We need to stop being myopic in our definition of what and where we can do our science. **Heliophysics at NASA – and thus the broader solar and space physics community - should be unbound from its current confinement to the applied science of space weather and the Sun-Earth connection and be made more universally relevant by expanding to studies of the fundamental nature of space plasma physics across the solar system and greater cosmos.** A trend in this direction has already begun. Parker Solar Probe has produced exciting results from its encounters with Venus. Heliophysics is funding the ESCAPADE mission to Mars and has assumed responsibility of the Radiation Assessment Detector (RAD) experiment on the Curiosity rover. The National Science Foundation (NSF) has encouraged comparative magnetospheric studies in its most recent call for the Geospace Environment Modeling (GEM) program. NASA Heliophysics has also funded mission concept studies to explore the local interstellar medium, the radiation belts of Jupiter, and the magnetosphere of Uranus. Perhaps an even more dramatic rebranding could be considered - perhaps the "Heliophysics Division" moniker could be changed to something more expansive and easily accessible, such as the "Solar and Space Physics Division"[1].

**We need to do more to introduce a sense of wonder and exploration into our science and our missions.** The lack of growth that Heliophysics has seen is in part a result of the lack of coherency in the Heliophysics community on how to define and "sell" our science. Heliophysics missions also often tend to lack a strong inspirational tagline that can compete with those of other SMD Divisions – e.g., planetary missions "boldly going" to new unexplored targets and astrophysical missions "unveiling the mysteries of the universe". A notable recent exception and resounding success would be the Parker Solar Probe mission – Heliophysics' most recent Living With a Star (LWS) mission "to touch the Sun," which had an incredible public presence leveraging its pioneering exploration of the solar corona and the engineering marvels required to achieve it.

---

[1] A "Space Physics Division" existed in NASA's Office of Space Science from 1987 to 1995.

**We need to come together as a community to advance our science by envisioning, prioritizing, and supporting with a unified voice a set of bold new missions that target compelling science questions– even if they do not explore "our" aspect of Heliophysics science – because these large missions can be the key to galvanizing the public and policymakers to support the overall Heliophysics program.** We need to adopt a "rising tide raises all boats" mentality. Larger-scale strategic missions are the "long poles" that support the tent under which all of a NASA SMD Division's research lives. Other NASA SMD Divisions have shown how major, flagship-scale missions have driven growth in their funding. It is by conceiving and promoting these major missions that we will be able to drive growth in the Heliophysics budget that will benefit the entirety of the community.

A potential vehicle to both enable these exciting new missions and to leverage growth in the Heliophysics Division's overall budget would be to redesign the current Heliophysics Division program. Prior to the previous (2013) Decadal Survey, the Solar-Terrestrial Probes (STP) program was essentially the Heliophysics equivalent of a large-scale strategic (i.e., "flagship") mission line. That was changed in the 2013 Decadal Survey when it was recommended that the STP missions be transitioned to a Principal Investigator-led line capped at $520M. Parker Solar Probe – which was already underway during the last Decadal Survey – was a large-scale mission under the LWS program, but the programmatic distinctions between LWS and STP – i.e., where fundamental science end and application begins – remain extremely confusing

To address this, **a specific suggestion is to rename the "Solar-Terrestrial Probes" program to "Heliophysics Probes" and have it support "flagship"-class missions of >$1B**; this would both create a dedicated program line for large-scale strategic missions and also address the overly narrow focus of the name of the program itself, allowing for much broader scientific scope. A new approach to a "flagship line" may be necessary to accommodate the breadth of physical regions covered by Heliophysics and the need to often embed or send spacecraft to sample environments in situ, as opposed to Astrophysics' or Earth Science's ability to use a single monolithic flagship mission as a community resource. To this end, perhaps this new "Heliophysics Probes" program could recommend a network of specifically and intentionally designed missions that complement each other and together work to explore solar and space physics processes across the solar system. For example, this "Heliospheric Extended Research Observatory (HERO)" Program could incorporate large-scale, strategic missions each specifically focusing on solar,

inner heliospheric, magnetospheric, ionosphere-thermosphere-mesosphere, and outer heliospheric objectives that *intentionally fit together* under a broader goal to enable end-to-end, systems-scale understanding of the coupled and interconnected heliospheric system-of-systems. Meanwhile, the LWS program would be transitioned from a "mission line" to an "objectives-based line". This new program would establish a set of measurement objectives necessary to address our fundamental scientific understanding of living with a star (not operational space weather needs), which could be filled by various approaches and/or targeted missions of various scales from rideshare payloads to missions on CubeSats or larger spacecraft. This would then leave a potentially large gap in the Heliophysics portfolio for missions between the Medium-scale Explorers (MIDEX) and the new >$1B Heliophysics Probes. This could be addressed by creating a new "Heliophysics Discovery" program line that would mirror the Discovery Program in the NASA Planetary Science Division: an open competition for missions with budgets under $500M.

*Summary*

The Heliophysics community:
- has needs to clearly and succinctly define our science in a compelling way that answers "*What is Heliophysics?*";
- has developed an overreliance on space weather to define our field, thus narrowing the scope of what solar and space physics is and diminishing its fundamental importance;
- should unapologetically promote their fundamental and practical research and the genuinely universal scope of Heliophysics to the public and policymakers;
- looking forward to the next decade, needs to be bold and unabashed in our definition of Heliophysics and its big questions;
- should be unbound from its current confinement to the Sun-Earth connection and expanded to studies of space plasma physics across the solar system and beyond;
- must come together to envision, prioritize, and support a set of bold new missions; and
- should advocate for restructuring the Heliophysics program: a new "Heliophysics Probes" line to support (not PI-led) "flagship"-class missions of >$1B; restructuring the LWS Program as an "objectives-based line; creation of a new "Heliophysics Discovery" Program, an open competition for missions with budgets under $500M; along with the current Explorers Program and Heliophysics Research lines.
  - An envisioned "Heliospheric Extended Research Observatory (HERO)" Program could incorporate large-scale, strategic missions each specifically focusing on solar, inner heliospheric, magnetospheric, ionosphere-thermosphere-mesosphere, and outer heliospheric objectives that *intentionally fit together* under a broader goal to enable end-to-end understanding of the coupled heliospheric system.